\documentclass[prl,twocolumn,showpacs,showkeys,preprintnumbers,floatfix,nofootinbib]{revtex4-1}
\pdfoutput=1
\usepackage{amsmath, amssymb, amsthm, color,latexsym,graphicx,array,multirow}
\usepackage{dcolumn} 
\usepackage[caption=false, labelformat=empty]{subfig} %

\newcommand{\gev}{\text{GeV}}

\newcommand{\kt}{k_T}
\newcommand{\CA}{\text{C/A}}
\newcommand{\Akt}{\text{anti-}k_{T} }

\newcommand{\Eq}[1]{Eq.~\eqref{#1}}
\newcommand{\Fig}[1]{Fig.~\ref{#1}}

\begin{document}

\title{Jets and Photons}

\author{Stephen D.~Ellis}
\affiliation{Department of Physics, University of Washington, Seattle, WA 98195} 
\author{Tuhin S.~Roy}
\affiliation{Department of Physics, University of Washington, Seattle, WA 98195} 
\author{Jakub Scholtz}
\affiliation{Department of Physics, University of Washington, Seattle, WA 98195} 

\date{\today}

\begin{abstract}
 
This Letter applies the concept of `jets',  as constructed from calorimeter cell $4$-vectors, to jets composed (primarily) of photons (or leptons). Thus jets become a superset of both traditional objects such as QCD-jets, photons, and electrons, and more unconventional objects such as photon-jets and electron-jets, defined as collinear photons and electrons, respectively. Since standard objects such as single photons become a subset of jets in this approach, standard jet substructure techniques are incorporated into the photon finder toolbox. We demonstrate that, for a single photon identification efficiency of $80\%$ or above, the use of jet substructure techniques reduces the number of QCD-jets faking photons by factors of $2.5$ to $4$. Depending on the topology of the photon-jets, the substructure variables reduce the number of photon-jets faking single photons by factors of $10$ to $10^3$ at a single photon identification efficiency of $80\%$. 

\end{abstract}

\maketitle

The final states in a  collider experiment are characterized in terms of a handful of objects. The detectors are designed to detect photons, electrons, muons, and a small number of hadrons (mostly charged pions) because these are the only stable objects in the Standard Model, apart from neutrinos that exit undetected.  
Converting and associating the various signals in different parts of the detector to the familiar physics objects is a non-trivial challenge. In this Letter, we discuss the class of objects that dominantly deposit their energy in the high density materials of the calorimeter component of a detector. Photons and electrons are absorbed in the inner part of the calorimeter (the electromagnetic calorimeter or ECal) while the hadrons are absorbed in the outer part (the hadronic calorimeter or HCal).

It is important to note that the content of the final state evolves as it moves out from the interaction point. At very short times and distances (typically less than $10^{-15}~\text{m}$) the final state consists of leptons, photons, and partons. 
The color charged partons rapidly radiate more (largely collinear) partons forming showers of partons, and subsequently get organized into showers of color neutral hadrons.  Most of these hadrons decay before or within the detector into lower mass hadrons, photons, and leptons. Consequently, photons and leptons can be part of a QCD-shower.  Typically, energy deposits in the ECal are identified as {\it isolated} photons/electrons (i.e., not associated with a QCD shower) if they satisfy various isolation and shower-shape criteria.  The remainder of the energy deposited in the ECal and HCal is clustered together using a specific `jet algorithm', to construct jets.  A small fraction of these jets are tagged as arising from the hadronic decays of $\tau$ leptons, based on another set of isolation and shape variables, and are removed from the list of jets. An event, therefore, is primarily classified in terms of the number of isolated photon/leptons and jets observed, along with their kinematic properties.

As suggested above, jets are often interpreted as the experimental `footprints' of {\it single} energetic partons produced in hard scattering events.  A more sophisticated analysis reveals that such associations are naive:  the jets identified using a typical jet algorithm will always contain contributions from the color-connected (but kinematically uncorrelated) soft component of the same hard collision (the `underlying event' or UE) and (at high luminosity) from truly uncorrelated but essentially simultaneous collisions of other beam particles (`Pile-Up' or PU). Moreover, jets often contain the showers arising from more than one energetic parton.  The jet-parton mapping breaks down further when we consider photon-jets~\cite{Dobrescu:2000jt, Toro:2012sv, Draper:2012xt} or electron-jets~\cite{Ruderman:2009tj, Cheung:2009su,Falkowski:2010cm,Falkowski:2010gv} that fail to be identified as isolated photons or electrons and are accepted as jets. More importantly, if the photons inside the photon-jets are highly collimated, they may fake single photons.  If the rate at which these photon-jets pass the detector definition of photons is large, the measurements performed interpreting the detected calorimeter objects as single photons become unreliable.

The issues raised above are extremely important in the context of Higgs physics. There are new physics scenarios where the Higgs particle decays into photon-jets at a rate comparable to, or even larger than, its decay to single photons~\cite{Dobrescu:2000jt, Draper:2012xt}.  The precise measurement of the $h \rightarrow \gamma \gamma$ rate requires a clean separation of photons from photon-jets (as well as from QCD-jets).  At the same time, we need a procedure that clearly distinguishes the photon-jets (and also electron-jets) from QCD-jets, since these photon-jet decay modes for the Higgs can provide signatures of physics Beyond the Standard Model~\cite{Schabinger:2005ei,Patt:2006fw,Strassler:2006im}. In other words, it is essential to extend the list of detectable/identifiable objects to include photon-jets and electron-jets with reliable separation from single photons/electrons and from QCD-jets.

In this Letter we propose such a formalism. The key ingredient is that we take `jets', defined as the output of a standard (IR safe) jet algorithm, to be the common construct for \emph{all} physics objects that deposit energy in the calorimeters.  A subsequent analysis of these jets, especially using recently defined jet substructure variables~\cite{Seymour:1993mx, Brooijmans:1077731,Butterworth:2007ke, Butterworth:2008iy, Thaler:2008ju, Kaplan:2008ie}, allows the jets to be identified and associated with the appropriate physics objects.

Note that we draw a clear distinction between the terminology of `jets' and `QCD-jets' in this Letter. We define `jets' as the output of jet algorithms such as $\Akt$~\cite{Cacciari:2008gp}, $\kt$~\cite{Catani:1993hr, Ellis:1993tq}, or $\CA$~\cite{Dokshitzer:1997in, Wobisch:1998wt, Wobisch:2000dk}, which, in some instances, may have nothing to do with the usual QCD partons.  A jet, therefore, is a  generic concept that is defined in terms of the energy deposited in calorimeter cells and identified by a jet algorithm.  With this definition a QCD-jet is simply a special kind of jet, as is a photon/electron or any other conventional/unconventional calorimeter based object.

To distinguish jets of various kinds, we take a multivariate approach. We use a set of observables to train a boosted decision tree (BDT)~\cite{BDT} to optimize separation. The conventional variables that are often used to distinguish a photon/electron from QCD-jets~\cite{Bayatian:942733, ATL-PHYS-PUB-2011-007} can be applied in our jet-based formalism without compromising their efficiency.  The additional power of our formalism arises from  including jet substructure variables.

Before proceeding, we summarize the advantages of using jets as the fundamental objects. First, jets provide a unifying language for \textit{all} calorimeter objects, which eliminates the previous need to use different constructions for QCD-jets and photons/electrons. Second, jet substructure based observables provide additional power for discriminating among the various kinds of jets. Finally, performing a jet substructure based analysis on objects such as single photons/electrons and also photon/electron-jets, means that grooming techniques (such as filtering~\cite{Butterworth:2008iy, Butterworth:2008sd, Butterworth:2008tr}, pruning~\cite{Ellis:2009su, Ellis:2009me}, trimming~\cite{Krohn:2009th}), developed mainly in the context of QCD-jets, can now be applied to these objects.  Such grooming serves to reduce contributions from the UE and PU~\cite{ATLAS-CONF-2012-065, CMS-PAS-EXO-11-095}.

The efficacy of the above approach will be demonstrated through explicit examples from Higgs physics.  We consider three kinds of events: events with $pp \rightarrow h+X \rightarrow \gamma \gamma +X$, events with  $pp \rightarrow h+X \rightarrow 2~\text{photon-jets}+X$,  and finally,  QCD dijet events.  These events provide us with samples of single photons (i.e., jets dominated by single photons), photon-jets (jets containing several energetic photons), and QCD-jets.  Here we concentrate our discussion on the extraction of single photon samples, minimizing the backgrounds due to QCD-jets as and photon-jets. The analysis based on conventional variables shows substantial separation between photons and QCD-jets, but fails to separate photons from photon-jets.  The jet substructure variables, when used along with the conventional variables, provide further separation between single photons and QCD-jets. This enhanced analysis can separate single photons from photon-jets,  photon-jets from QCD-jets, and even offers the possibility of determining details of any new physics scenario that leads to such photon-jets. In this Letter we show only the final results of the multivariable analyses and discuss photon-jets of just two particular topologies.  A more exhaustive study of the many different discriminating variables along with analyses comparing photon-jets of varied topologies  will be presented elsewhere~\cite{Ellis:2012zp}.

In the rest of the paper we present brief descriptions of the discriminating variables and simulation details,  followed by a summary of our results. These results demonstrate how well single photons, photon-jets and QCD-jets can be differentiated from each other and also quantify the role played by the jet substructure variables.

We use two conventional variables that play essential roles in separating photon/electrons from QCD-jets. 

\noindent \textbf{Hadronic Energy Fraction ($\theta_J$):} 
The most powerful observable for discriminating a photon from a QCD-jet stems from the fact that a QCD-jet almost always deposits some energy in the HCal. A  QCD-jet consists of mostly pions and, on average, 2/3 of these pions are charged. The charged pions lose most of their energy in the HCal. The HCal isolation criterion exploits the feature that, for a photon to be isolated, the energy deposited in the HCal  (within a cone about the direction of the photon and of a given size) must be significantly smaller than the energy of the photon-candidate itself.  It is straightforward to implement the isolation criterion in terms of the `Hadronic Energy Fraction', defined as the fraction of the total jet energy deposited in the HCal, $\theta_J = E_{J,\text{HCal}}/E_{J}$.  In the analysis described below \textit{all} included jets are required to pass a cut $\theta_J \leq 0.25$, which eliminates~\cite{Ellis:2012zp} about $98\%$ of the QCD-jets but keeps about $94\%$ of the single photons and photon-jets.

\noindent \textbf{Number of Hard Tracks ($\nu_J$):}  
Charged particles leave tracks in the Tracker portion of the detector, where they bend due to the presence of a magnetic field allowing a measurement of their momenta and charges. We count the number of charged particles with $p_T>2~\gev$ present in the jet, which we label $\nu_J$.  This variable can discriminate photons and photon-jets (characterized by $\nu_J=0$ if photons do not convert) from QCD-jets, which often contain a large number of charged pions. Operationally, we define a charged particle to be ``present in the jet", if a light-like and arbitrarily soft four-vector, having the same $\eta,\phi$ as the charged particle, is clustered into the jet when we apply the jet 
algorithm to the original calorimeter cell four-vectors plus these new soft four-vectors.

We do not include a `calorimetric isolation variable', defined as the fraction of energy deposited in the outer annulus of an inner cone for a given jet.  Independent of the radius of the inner cone, using a calorimetric isolation variable along with $\theta_J$ and $\nu_J$, further reduces the QCD-jet fake rate at most by order $10$-$20\%$, and fails to reduce photon-jets faking photons. Often observables based on shower-evolution or particle-flow inside the detector are used to discriminate photons/electrons from QCD-jets.  While we do not include these variables in the current work,  we do not foresee any difficulty in using such variables in the context of the jet analysis described here.

The rest of the variables we use are constructed using exclusive subjets of jets. The (calorimeter cell) constituents of a given jet, identified by the jet finding algorithm, are (re)clustered using the $\CA$ or $\kt$ algorithm until there remain exactly $N$ four-vectors, i.e., the reclustering is halted by the constraint on $N$, not the algorithm parameter. These are the $N$ exclusive ($\CA$ or $\kt$)-subjets of the given jet.

\noindent \textbf{Nsubjettiness ($\tau_N$):}   
$N$-subjettiness, as introduced in Ref.~\cite{Thaler:2010tr, Thaler:2011gf, Stewart:2010tn}, provides a simple way to effectively count the number of energetic subjets within a given jet, and hence to discriminate among jets with varied energy flows.  
For a given jet and its $N$ exclusive $\kt$-subjets we evaluate the $N$-subjettiness using the expression~\cite{Thaler:2010tr}.
\begin{equation}
	\label{eq:tau}
	\tau_N = \frac{ \sum_k  p_{T_{k}}  \times \text{min} \bigl \{ \Delta R_{1,k}, \Delta R_{1,k} , 
				\cdots, \Delta R_{N,k}  \bigr \} } { \sum_k  p_{T_{k}} \times R}  \; ,
\end{equation}
where $k$ runs over all the constituents of a jet,  $\Delta R_{l,k} = \sqrt{(\Delta \eta_{l,k})^2 + (\Delta \phi_{l,k})^2}$, is the angular distance between the $l$-th subjet and the $k$-th constituent of the jet, and   $R$ is the characteristic jet radius used in the jet clustering algorithm.  For a jet with $N_0$ actual energetic subjets, the value of $\tau_N$ will be substantially larger for $N < N_0$ than for $N \geq N_0$, allowing a `measurement' of $N_0$.

\noindent \textbf{Subjet distributions in a `filtered' jet:} 
We consider $5$ exclusive subjets for a given jet and, out of these, only use the $3$  largest $p_T$ subjets to construct the observables defined in \Eq{eq:substructure}. Note that, by discarding the $2$ softest subjets, we have performed a version of `grooming' typically labeled filtering.  This ensures that our results are relatively insensitive to the effects of the UE and PU. We use the following four variables to quantify how the leading subjets are distributed inside the jet. 
\begin{equation}
    \label{eq:substructure}
    \begin{split}
	\lambda_J \  = \ \log \Bigl( 1 - \frac{ p_{T_L}} {p_{T_J}} \Bigr)\; ,     \qquad   &
 	\epsilon_J \  = \  \frac{1}{E_J^2}  \sum_{i > j }  E_{i} E_{j} \; ,\\
	 \rho_J  = \frac{1}{R} \sum_{i > j}  \Delta R_{i,j} \; ,     \qquad  &
	\delta_J  = \frac{1}{A_J} \sum_{i} A_i   \; .
	\end{split}
  \end{equation}
In these equations we use the following definitions: $p_{T_J}$, $E_J$, $A_J$ are the transverse momentum, energy, and active area~\cite{Cacciari:2008gn} of the given jet;  $ E_i$ and $A_i$  are the energy and active area of the $i$-th subjet; $p_{T_L}$ is the $p_T$ of the leading subjet; and $\Delta R_{i,j}$ is the angular distance between the $i$ and $j$-th subjet.  The variable $\lambda_J$ characterizes the fraction of jet $p_T$ carried by the leading subjet.  The variable  $\epsilon_J$ encodes information about how the jet's energy is shared among the subjets. The geometric observable $\rho_J$ carries information on the spatial distribution of subjets inside the jet, while $\delta_J$ characterizes the `cleanliness' of the jet. In the spirit of Ref.~\cite{Ellis:2012sn}, we use both $\kt$ and $\CA$ subjets to calculate the variables in \Eq{eq:substructure}. Also, these observables depend on how we select the subjets. We find that the choice of ``$3$ out of $5$" for filtering to be optimal for separating photons from photon-jets with a range of photon-jet topologies.

In order to minimize the background fake rate for a given signal acceptance, we include all the variables described above in BDTs as implemented in the Toolkit for Multivariate Analysis~\cite{Hocker:2007ht}.  Given a signal and background we construct three separate BDTs, each optimized using the following three sets of variables: 
\begin{equation}
	\label{eq:variable_sets}
	\begin{split}
		D \  \equiv \ &  \Bigl\{ \log \theta_J, \nu_J,   \log \tau_1, \frac{\tau_2}{\tau_1}, 
					\frac{\tau_3}{\tau_2}, \frac{\tau_4}{\tau_3},   \\ &  \quad
								\bigl( \lambda_J,  \epsilon_J, \rho_J,  \delta_J \bigr)  \bigl|_{\text{C/A}},  	
			  				\bigl( \lambda_J,  \epsilon_J, \rho_J \bigr)  \bigl|_{k_T}  \Big\}   \; ,
				 \\
		D_{\text{C}} 	\  \equiv \  &  	\Big\{\log \theta_J, \nu_J \Big\}  \; , \quad \text{and} \quad 
		D_{\text{S}} 	\  \equiv \   D - D_{\text{C}} 		\; ,	
	\end{split}
\end{equation}
where the subscript $\CA$ or $\kt$  means that the observables are calculated using $\CA$ or $\kt$ subjets. The sets $D_{\text{C}} $ and $D_{\text{S}}$  consist of conventional and jet substructure variables respectively.  $D$ is the set of all variables.

We generate all events with Pythia~$8$~\cite{Sjostrand:2007gs}. For photon-jets we set up a model in MadGraph~$5$~\cite{Alwall:2011uj}, where the Higgs particle decays to a pair of new light scalars ($n_1$) of mass $m_1$. We simulate photon-jets with two photons by allowing the decay $n_1 \rightarrow \gamma \gamma $.  For photon-jets with four photons we force the $n_1$ to decay via $n_1 \rightarrow n_2\left(\rightarrow \gamma \gamma\right) n_2\left(\rightarrow \gamma \gamma\right)$, where $n_2$ is a second scalar with mass $m_2$. In this work we set the Higgs mass at $120~\gev$; $m_1 = 1~\gev$ to simulate $2$~photon photon-jets; and 
$m_1=5~\gev, m_2 = 1~\gev$ for photon-jets with $4$~photons. These choices of parameters ensure that  the decay products of  the $n_1$ are highly collimated and are usually contained in a single jet.   We use the default scheme for the UE as implemented in Pythia~$8$ to simulate an appropriately busy hadronic environment. 

To simulate a (reasonably) realistic calorimeter the photons, electrons, and hadrons in a Pythia event are grouped into ECal cells of size $0.025 \times 0.025$, and HCal cells of size $0.1 \times 0.1$ in the ($\eta$-$\phi$) plane.  We incorporate aspects of transverse showering for photons inside the ECal as well as calorimeter energy smearing for both the ECal and the HCal.  We also simulate the conversion for photons into $e^+e^-$ pairs. Note, however, that we do not  include effects of a magnetic field inside the detector.  Using the total energy deposited in a cell and its $(\eta, \phi)$ coordinates we construct light-like momentum four-vectors for each cell. These four-vectors, corresponding to the ECal and HCal cells, contribute to the analysis only if they pass the energy threshold of  $0.1~\gev$ (ECal) and $0.5~\gev$ (HCal). We use the $\Akt$ algorithm  as implemented in FastJet~\cite{Cacciari:2011ma,*Cacciari:2005hq} to cluster the calorimeter cells into jets with $R = 0.4$. Only the leading $p_T$ jet, with $p_T > 50~\gev$, from each event is used in the analysis.  

\begin{figure}[t]
	\subfloat[]{\includegraphics[width=0.23\textwidth]{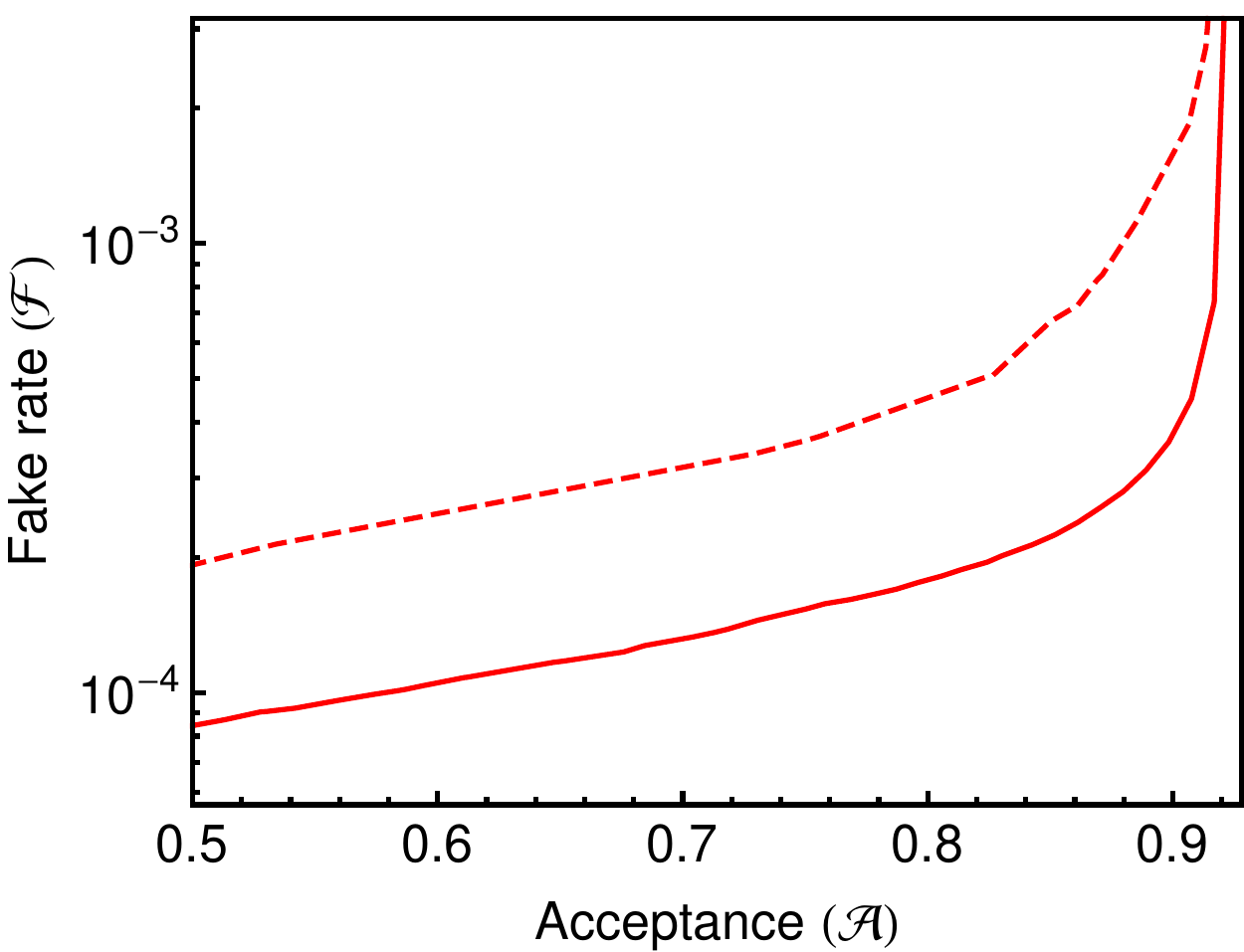}} \hspace{1.mm}
	\subfloat[]{\includegraphics[width=0.23\textwidth]{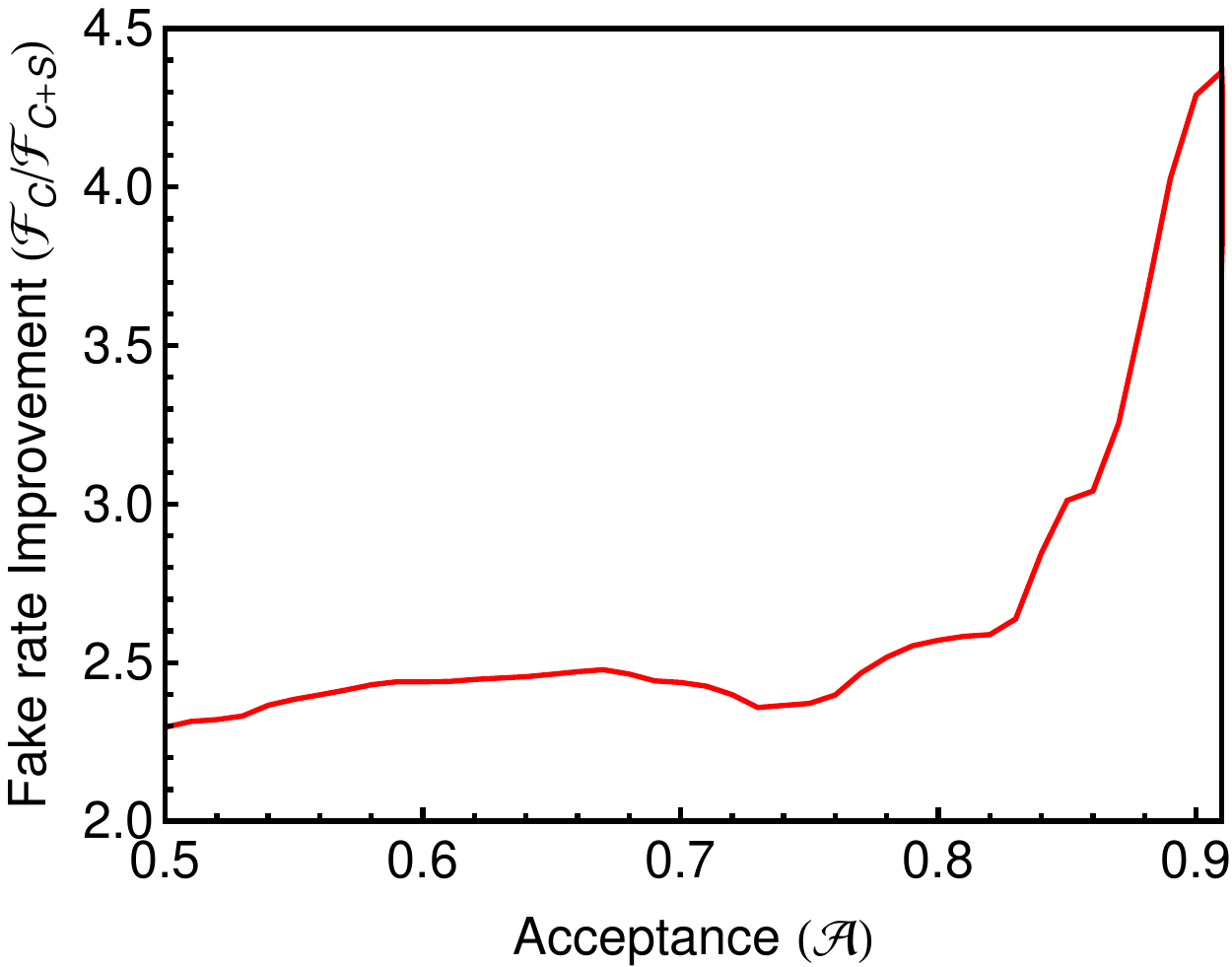}} \\
	\vspace{-10.5 mm}
	\subfloat[]{\includegraphics[width=0.23\textwidth]{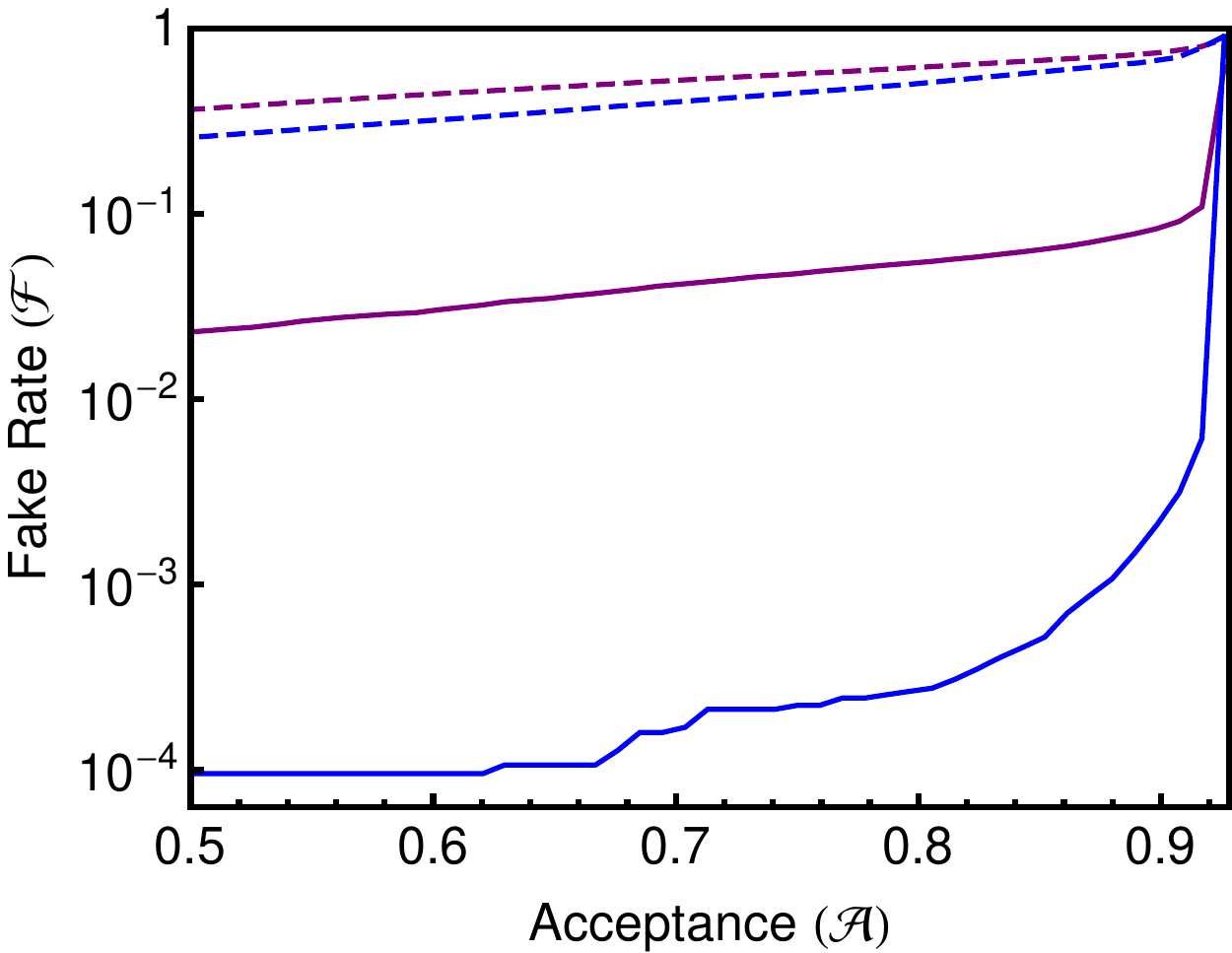}} \hspace{1.mm}
	\subfloat[]{\includegraphics[width=0.23\textwidth]{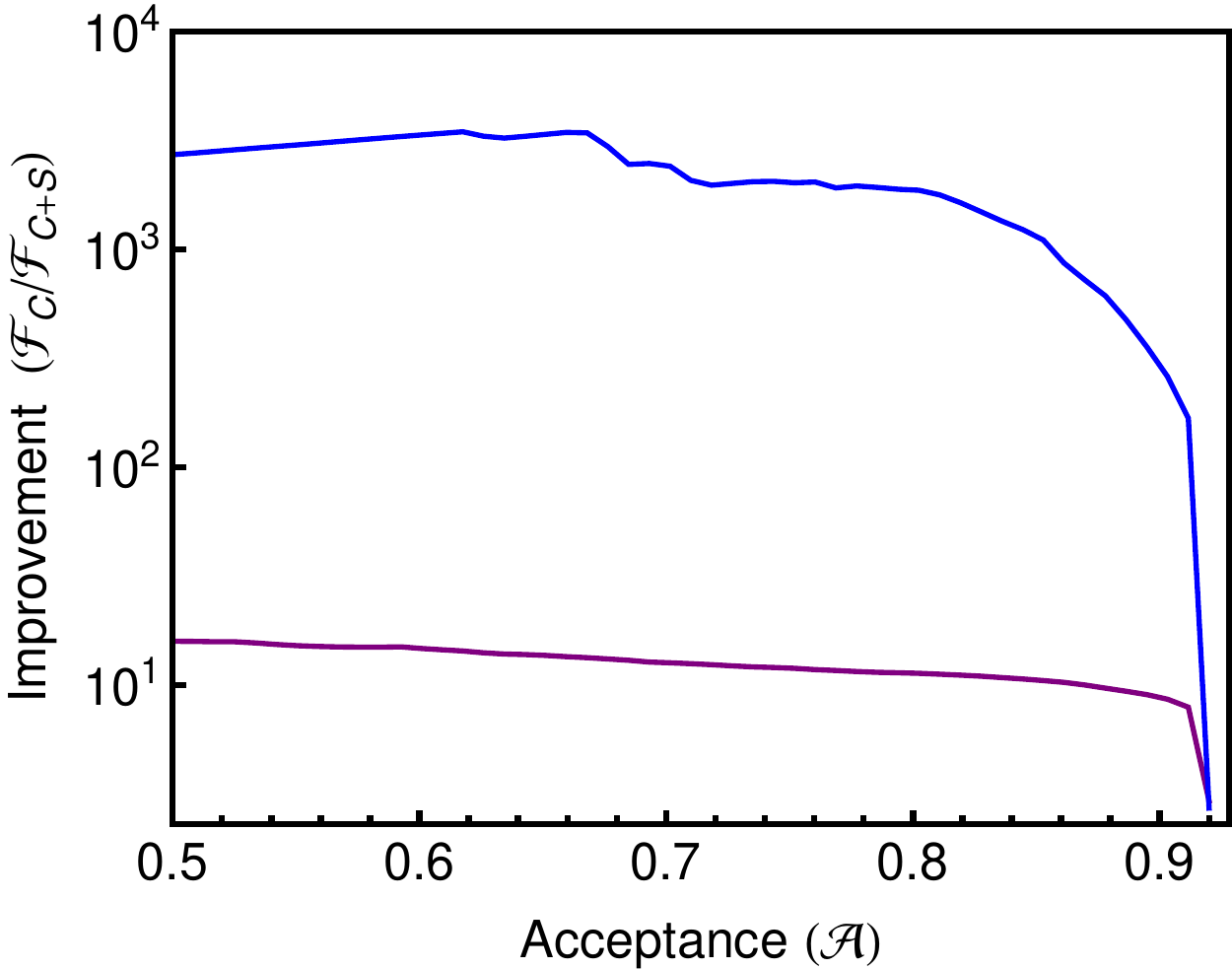}}
		\vspace{-5 mm}
\caption{\label{fig:PvsBkg} The plots in the left column show the background fake rate $\mathcal{F}$ versus the single photon acceptance $\mathcal{A}$,  where the solid (dotted) lines corresponds to  BDTs using the full $D$ variable set ($D_\text{C}$ set only).  The right panel indicates the extra suppression of the fake rate arising from including the jet substructure variables. In these figures, the red, maroon, and blue colored curves designate the cases when the background is due to QCD-jets, $2$ photon photon-jets, and  $4$ photon photon-jets, respectively. }
\vspace{-5mm}
\end{figure}

In this Letter we report our results for three separate questions.  $(i)$ With single photons treated as the signal, we determine how well we can reduce the rate at which QCD-jets fake single photons. $(ii)$  We perform the same analysis treating photon-jets as the background to single photons. $(iii)$ Finally, we seek to separate single photons from photon-jets, while, at the same time, attempting to keep QCD-jets from faking either of these. 

In \Fig{fig:PvsBkg} (left panels), we display the results for the fake rate ($\mathcal{F}$) versus the acceptance ($\mathcal{A}$) for single photons treating either QCD-jets (top-row) or photon-jets (bottom-row) as the background.  
In the right panels we characterize the improvement in separation allowed by including the jet substructure variables. For a given signal acceptance, we define the improvement to be the ratio of fake rates $\mathcal{F}_\text{C}/\mathcal{F}_\text{C+S}$, where $\mathcal{F}_\text{C}$ and $\mathcal{F}_\text{C+S}$ are the fake rates if the BDTs are optimized using the variables in $D_{\text{C}}$ and $D$, respectively.

The top panel in \Fig{fig:PvsBkg} shows that the conventional variables already provide significant separation between single photons and QCD-jets. The substructure variables reduce the fake rate by an additional factor of $2.5$ for a single photon acceptance of $80\%$, resulting in a total fake rate of about $1$ in $10^4$.  For larger acceptance values the fake rate increases, but the improvement due the substructure variables also increases to a value above $4$.  The separation of single photons from the photon-jets, on the other hand, are entirely due to the jet substructure variables as indicated in the bottom panel of \Fig{fig:PvsBkg}. Comparing the $2$ photon photon-jets (maroon) case with the $4$ photon photon-jets (blue) indicates that single photons can be separated more efficiently from the $4$ photon photon-jet background than from photon-jets with $2$ photons. Having multiple photons inside the jet ensures that the energy in the jet is distributed in multiple subjets imparting more substructure to the jet.
We find that for single photon acceptances over $80\%$, we can obtain fake rates as low as $2\times 10^{-4}$ for QCD-jets, $0.05$ for $2$ photon photon-jets, and $3\times 10^{-4}$ for $4$ photon photon-jets.Ó

\begin{figure}[t]
	\centering
	\subfloat[]	{\includegraphics[width=0.23\textwidth]{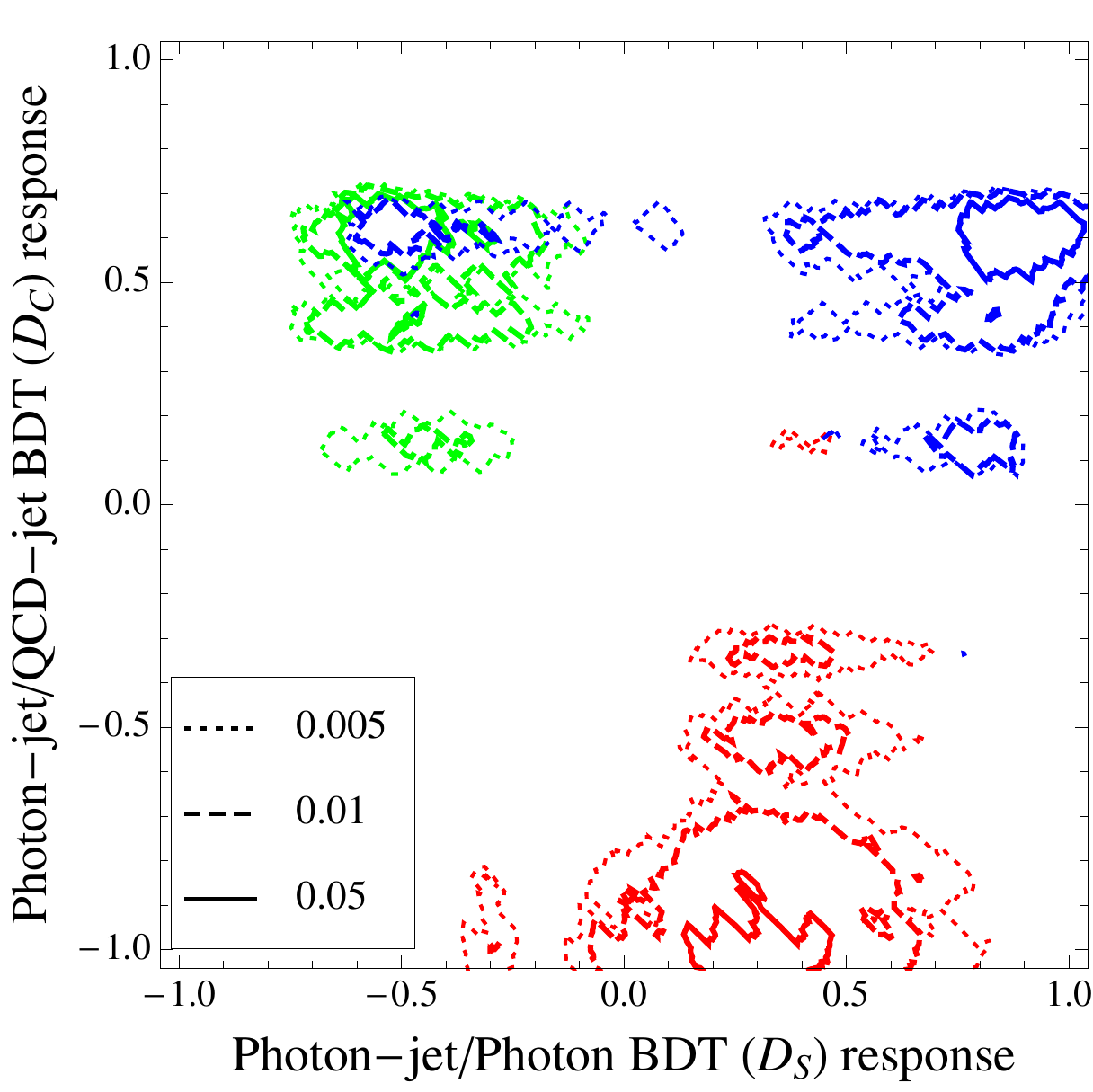} } \hspace{-1.mm}
	\subfloat[]{\includegraphics[width=0.23\textwidth]{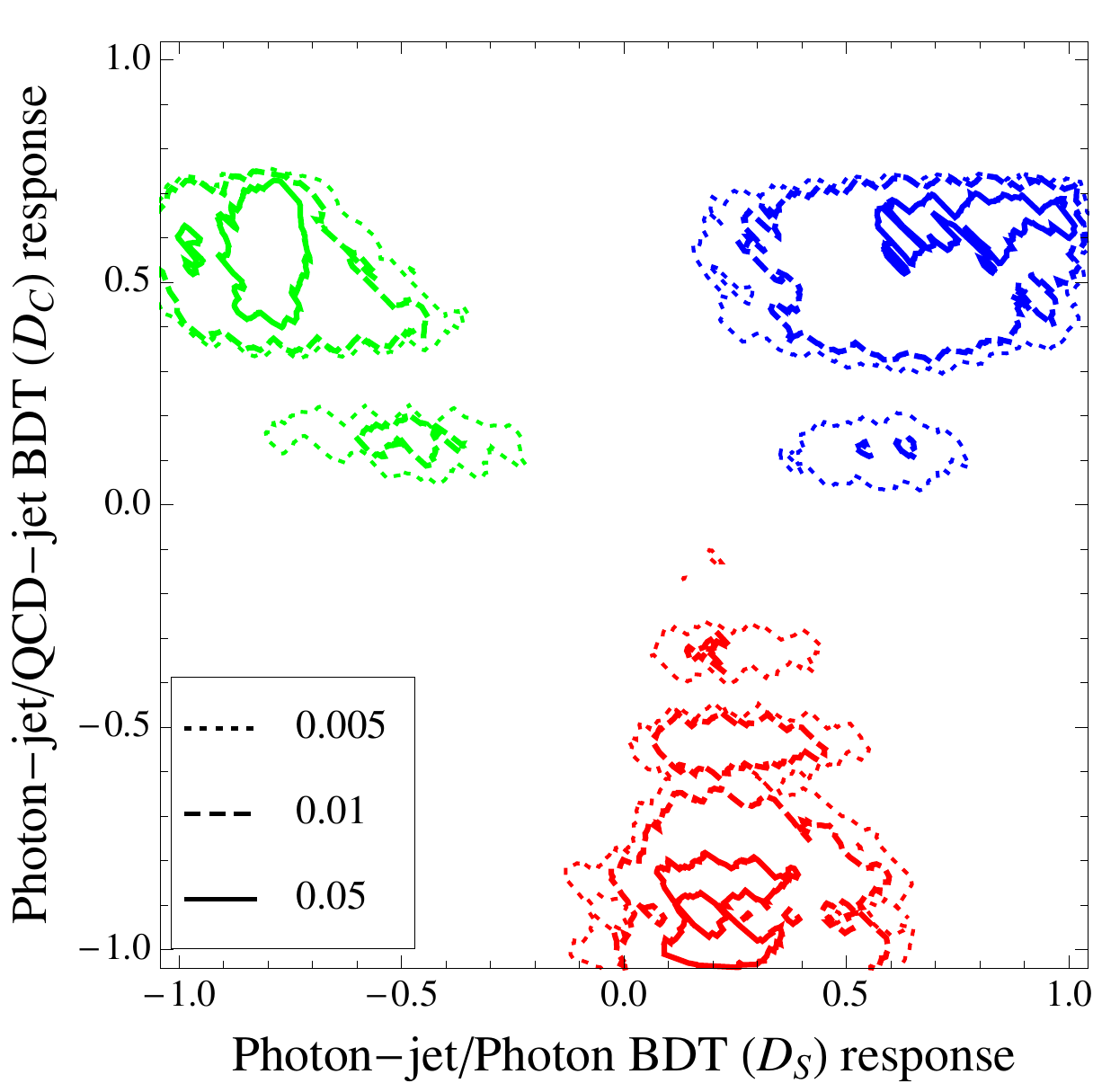} }	
		\vspace{-6 mm}
	\caption{\label{fig:Density_plots} The BDT responses for QCD-jets (red), photons (green) and photon-jets (blue). The left  (right) panel shows photon-jets containing $2$ photons ($4$ photons).   }
\vspace{-7mm}	
\end{figure}

Figure~\ref{fig:Density_plots} displays an example of three-way separation between single photons, photon-jets, and QCD-jets using two BDTs.  The first BDT is optimized to separate photon-jets from QCD-jets employing only the conventional ($D_\text{C}$) variables (and its response is plotted on the vertical axis). The second BDT is trained to separate photon-jets from single photons using only the jet substructure ($D_\text{S}$) variables (and its response is plotted on the horizontal axis).   By construction the upper left corner is primarily single photons, the upper right is primarily photon-jets, and QCD-jets tend to lie along the bottom axis.  The left (right) panel corresponds to photon-jets with $2$ ($4$) photons.  In the two-dimensional space of the responses of these two BDTs,  the numerical values associated with a given contour corresponds to the relative probability to find a calorimeter object in a cell of size $0.1\times 0.1$ in BDT response units, which range from -1 (background-like) to +1 (signal-like).  As indicated in Fig.~\ref{fig:Density_plots},  separating photons from $2$ photon photon-jets remains challenging. A small fraction of the 2-photon photon-jet sample (of order few $\%$), represented by the dashed blue contours in the upper-left corner, constitute an irreducible background to photons. A much cleaner separation (for photon vs.~photon-jets) is observed for $4$ photon photon-jets.

In this work we have demonstrated the efficacy of using jet based techniques, including jet substructure variables, to analyze and identify the full class of objects constructed from the energy deposited in calorimeter cells.  This class includes not only the familiar single photons and QCD-jets, but also the potentially very interesting (at the LHC) photon-jets (and lepton-jets).  This approach not only has the advantage of defining a universal language for all such objects, but also enhances the possible analyses by allowing the inclusion of recent advances in jet substructure technology.  Previous efforts to distinguish these objects~\cite{ATLAS-CONF-2012-079} have largely used variables that are constructed in the spirit of substructure techniques, but treating everything as a jet allows a much more direct employment of jet substructure variables and analyses.  As we have shown, these can be powerful tools for identifying both single photons and photon-jets, separating them from QCD-jets and from each other.
  
\section*{Acknowledgments}
The authors would like to acknowledge stimulating conversations with  Henry Lubatti and Gordon Watts regarding this work. We especially thank Henry Lubatti for his careful reading of the manuscript.  TSR would like to thank the hospitality of CERN, where part of the work was completed. The work of SDE, TSR and JS was supported, in part, by the US Department of Energy under contract numbers DE-FGO2-96ER40956. JS would also like to acknowledge partial support from a DOE High Energy Physics Graduate Theory Fellowship.  Finally we acknowledge the importance of the computer time we used on the UW TEV cluster, which is supported by the US National Science Foundation grant number ARRA-NSF-0959141.


\end{document}